\documentstyle[12pt]{article}
\begin{document}

\centerline{\large SUPERNARROW SPECTRAL PEAKS AND HIGH}
\centerline{\large FREQUENCY STOCHASTIC RESONANCE IN}
\centerline{\large SYSTEMS WITH COEXISTING PERIODIC ATTRACTORS}

\vskip 0.3truecm
\centerline {M I Dykman}
\centerline {Department of Physics, Stanford University, Stanford,
CA 94305, USA.}

\vskip 0.3truecm
\centerline {D G Luchinsky\footnote{Permanent address: Research Institute
for Metrological Service, 117965 Moscow, Russia.}}
\centerline {School of Physics and Materials, Lancaster University,
Lancaster, LA1 4YB, UK.}

\vskip 0.3truecm
\centerline {R Mannella}
\centerline {Dipartimento di Fisica, Universit$\grave{\rm a}$ di Pisa, Piazza
Torricelli 2, 56100 Pisa, Italy.}

\vskip 0.3truecm
\centerline {P V E McClintock, N D Stein and
N G Stocks\footnote{Present address: Department of Engineering, University
of Warwick, Coventry, CV4 7AL, UK.}}
\centerline {School of Physics and Materials, Lancaster University,
Lancaster, LA1 4YB, UK.}

\vskip0.75truecm
\noindent
\underbar {Abstract}

The kinetics of a periodically driven nonlinear oscillator, bistable in a
nearly resonant field, has been investigated theoretically and through
analogue experiments.  An activation dependence of the
probabilities of fluctuational transitions between the coexisting attractors
has been observed, and the activation energies of the transitions have been
calculated and measured for a wide range of parameters.  The position of the
kinetic phase transition (KPT), at which the populations of the attractors are
equal, has been established.  A range of critical phenomena is shown to arise
in the vicinity of the KPT including, in particular, the appearance of a
supernarrow peak in the spectral density of the fluctuations, and the
occurrence of high-frequency stochastic resonance (HFSR).  The experimental
measurements of the transition probabilities, the KPT line, the multipeaked
spectral densities, the strength of the supernarrow spectral peak, and of
the HFSR effect are shown to be in good agreement with the theoretical
predictions.


\newpage
\centerline {I. INTRODUCTION}

The physical kinetics of bistable systems presents a number of intriguing
phenomena related, in particular, to the existence of substantially differing
characteristic relaxation times.  One (or several) of these, $t_r$,
characterise(s) the \lq\lq thermalization" of a system in the vicinity of
stable states, {\it i.e.} local relaxation and fluctuations {\it about}
these states.  The others characterise fluctuational transitions {\it between}
the stable states.  They are given by the reciprocal transition probabilities
$W^{-1}_{ij}$ (the indices {\it i, j} enumerate the stable states), and usually

\begin{equation}
W_{ij} t_r  \ll 1
\end{equation}

\noindent
The concept of bistability is meaningful, provided only that (1) is
fulfilled, in which case a system will spend most of the time
fluctuating about one or other of the stable states.  If the parameters
of the system pass through the range of bistability in a time much less
that $W^{-1}_{ij}$, the system will display hysteresis: it will tend to
remain within one or other of the stable states, depending on the prior
history.  For fixed system parameters, however, over times
$\sim W^{-1}_{ij}$ a stationary distribution over the stable
states is built up and the system \lq\lq forgets" which of the stable states
was occupied initially.

For thermal equilibrium systems the transition probabilities are usually
given by the Arrhenius law, $W \propto {\rm exp} (-E_a/T)$, where {\it T} is
the temperature and $E_a$ is the characteristic activation energy of the
transition.  In the case of a Brownian particle the quantity $E_a$ is
simply the depth of the potential well from which the particle escapes [1].
For nonequilibrium systems the calculation of the transition probabilities is
a nontrivial problem.  A rather general approach to its solution has been
proposed for dynamical systems driven by external Gaussian noise (see [2]
for a review).  In this case $W \propto {\rm exp} (-R/\alpha)$ where $\alpha$
is the noise intensity, while {\it R} is given by the solution of a certain
variational problem.

In the general case of a bistable system, the characteristic activation
energies $R_1$ and $R_2$ for the transitions 1 $\to$ 2 and 2 $\to$ 1 differ one
from
another. Consequently, for sufficiently weak noise, {\it i.e.} for small
$\alpha$ (when (1) is fulfilled and in addition {\it W} is much smaller than
the reciprocal correlation time of the noise), the probabilities $W_{12}$
and $W_{21}$ of the transitions 1 $\to$ 2 and 2 $\to$ 1 differ exponentially.
The latter is the
case also for the stationary populations $w_1$ and $w_2$ of the stable
states,

\begin{equation}
w_1 \,\, = \,\, \frac {W_{21}}{W_{12} + W_{21}}, \qquad
w_2 \,\, = \,\, \frac {W_{12}}{W_{12} + W_{21}}
\end{equation}

\noindent
For most parameter values of the system, the ratio $w_1/w_2$ is either
exponentially small (for $R_2 - R_1 \gg \alpha$) or large
(for $R_1 - R_2 \gg \alpha$) and the system occupies with an
overwhelming probability the states 2 or 1 respectively.  Only within a very
narrow range of parameters where $\vert R_1 - R_2 \vert \stackrel {<}{\sim}
\alpha$ are the populations $w_1$ and $w_2$ of the same order of magnitude.
In this range a kinetic phase transition occurs: as we discuss below, the
behaviour of a noise-driven dynamical system is to some extent analogous to
that of a thermodynamic system with coexisting phases ({\it e.g.}
liquid/vapour) within  the range of its first order phase transition, where
both phases are well manifested (with comparable molar volumes, for example).

A well-known signature of systems experiencing phase transitions is the
strong associated increase of fluctuations.  In the case of dynamical systems,
not only will there be small fluctuations about the stable states, but there
will also be large fluctuations related to transitions {\it between} the
states (the analogue of the fluctuational creation of macrobubbles in a
liquid/vapour system).  The characteristic time for such fluctuations is
obviously the reciprocal transition probability $W^{-1}$.  It is quite
natural to expect that these large and very slow fluctuations will give rise
to intense and extremely narrow (with a width $\sim W \propto {\rm exp}
(-R/\alpha)$) peaks in the susceptibility of the system and in the spectral
density of fluctuations (SDF) [3].  For a Brownian particle
fluctuating in a symmetric double-well potential (that is, exactly at the
phase-transition point, $w_1 = w_2$), the corresponding peak in the SDF at zero
frequency has already been observed [4].  The exponentially fast broadening of
this peak with increasing noise intensity gives rise [5] to low frequency
{\it stochastic resonance} [6-8] {\it i.e.} to the increase with increasing
noise of the signal and the signal/noise ratio in a system driven by a
low-frequency periodic force.

An important class of bistable systems is those displaying bistability
when driven by an intense periodic field, but which are monostable otherwise.
(Note that some such systems may also display multistability, and/or dynamical
chaos, when subjected to even stronger periodic fields.)  A variety of
them are investigated in, for example, nonlinear optics (in relation to
optical bistability, see [9]).  The different stable states here correspond to
periodic attractors with differing amplitudes, phases (and sometimes
frequencies) of constrained vibrations.  A well-known example [10] of such
a system is the nonlinear oscillator driven by a nearly resonant field.
This system is interesting not only as an archetypal model for the
investigation of a periodic-field induced bistability, but also because it
refers directly to a peculiar and interesting physical system, a relativistic
electron trapped in a magnetic field and excited by a resonant cyclotron
radiation [11].

In addition to its markedly nonequilibrium character, which provides a good
test for theories of fluctuational transitions in thermally nonequilibrium
systems, the model of a resonantly driven nonlinear oscillator also enables
one to investigate specific phenomena arising in systems with coexisting
states of forced vibration within the range of the kinetic phase
transition.  Since there is a special frequency in such systems,
namely the frequency of the external field $\omega_F$, the fluctuational
transitions between the stable states should modulate the response of the
system at frequency $\omega_F$: extremely tall and narrow spectral peaks near
$\omega_F$ are therefore to be expected, both in the susceptibility [2] and in
the SDF [12].  Such supernarrow spectral peaks have been observed in an
electronic analogue experiment [13].  Because the widths of such peaks
increase extremely rapidly (exponentially) with noise intensity, it is
to be anticipated that the signal/noise ratio for a signal at a frequency
close to $\omega_F$ will also increase with increasing noise intensity,
i.e. that there will be a manifestation of high frequency stochastic
resonance (HFSR).

In the present paper we present detailed results of our investigation  of the
features of the SDF in a periodically driven system, including the onset of
a supernarrow spectral peak in the region of the kinetic phase transition.
In Sec II below the theory of kinetic phenomena for a periodically
driven oscillator is given, including the results of a numerical calculation of
the \lq\lq activation energies" of the fluctuational transitions and explicit
expressions for the spectral density of fluctuations and for the generalized
susceptibilities.  In Sec III the experimental simulation of the oscillator
by an analogue electronic circuit is described.  In Sec IV the theoretical
and experimental results are compared with each other, and the new critical
phenomena, the onset of the supernarrow spectral peak and of the high-frequency
stochastic resonance, are discussed.  Sec V contains concluding remarks.

\newpage
\centerline {II. THEORY OF THE SPECTRAL DENSITY OF FLUCTUATIONS OF}
\centerline {AN OSCILLATOR BISTABLE IN A PERIODIC FIELD}

\noindent
\underbar {A. The Model}

In this section we explore the behaviour of a nonlinear oscillator subject
to the combined influences of a periodic field $F {\rm cos} \omega_Ft$ and a
weak random force {\it f(t)}.  The equation of motion of the particular
oscillator considered (single-well Duffing oscillator) is

\begin{equation}
\ddot q \,\, + \,\, 2\Gamma \dot q \,\, + \,\, \omega^2_0q \,\, + \,\,
\gamma q^3 \,\, = \,\, F{\rm cos} \omega_F t \,\, + \,\, f(t)
\end{equation}

\noindent
The oscillator is assumed underdamped and the periodic field nearly resonant

\begin{equation}
\Gamma, \,\, \vert \delta \omega \vert \ll \omega_F, \qquad \delta
\omega   =   \omega_F - \omega_0
\end{equation}

\noindent
A characteristic amplitude of vibration for which the oscillator will
obviously be strongly nonlinear, the nonlinear length $l_n$, is determined
by the condition that the nonlinear term $\gamma q^3$ in (3) should be
as large as the linear one, so that $l_n = (\omega_0^2/\vert \gamma
\vert)^{\frac{1}{2}}$.  If the amplitude $F$ of the periodic force is
sufficiently large that vibrations of amplitude $a \sim l_n$ are excited, then
the oscillator (3) in the absence of the random force $f (t)$ is known [14] to
display deterministic chaotic phenomena (see also [15]), with the boundaries
of the domains of attraction to various attractors often being fractal [16].

In the case of an underdamped oscillator, strong nonlinear effects
can also arise for much smaller values of $F$, for which the
vibration amplitudes $a$ are correspondingly much less than $l_n$
[10].  This is because the eigenfrequency of the vibrations of a nonlinear
oscillator depends on their amplitude, $\omega_{eff} \equiv \omega (a)$,
and it is the interrelation between the detuning of the field with
respect to the eigenfrequency and the damping, $\vert \omega_F - \omega (a)
\vert / \Gamma$, that determines the strength of the response.  For small
$\Gamma$, the latter ratio can vary markedly with $a$, even while $a \ll l_n$,
and it is this feature that can give rise to the coexistence of different
stable solutions for the amplitude $a$.  We may note that, for $a \ll l_n$,
the vibrational amplitudes at the overtones are $\sim a^3/l^2_n \ll a$,
and dynamical chaotic phenomena do not occur.  Because the noise intensities to
be considered here are relatively weak, the system seldom strays far from the
attractors, and practically never goes as far as $l_n$; the probability of
finding it there is exponentially small compared to the probabilities of
transitions.

Under conditions for which $\vert \gamma \vert \langle q^2 \rangle
\ll \omega_F$, the motion of the oscillator consists of relatively fast
oscillations with slowly varying amplitude and phase.  The characteristic
scale for these variations is determined by the friction coefficient $\Gamma$
and the detuning $\delta \omega$ of the field frequency $\omega_F$ with
respect to the oscillator eigenfrequency $\omega_0$: the characteristic
scale for the coordinate {\it q} is $\vert \gamma / \omega_F \delta \omega
\vert^{-\frac {1}{2}}$.  In describing the \lq\lq slow" motion it is
convenient, in the spirit of a standard averaging method (cf [17]) to
transform to the rotating frame.  We thus change from
$q, \dot q$ to the complex dimensionless envelopes $u, u^*$ and the
dimensionless time $\tau$,

$$q = \left( \frac {2\omega_F \vert \delta \omega \vert}{3 \vert \gamma \vert}
\right)^{\frac {1}{2}} \left( ue^{i\omega_Ft} + u^*e^{-i\omega_Ft}\right)$$

\begin{equation}
\dot q = i \omega_F \left( \frac {2 \omega_F \vert \delta \omega \vert}{3 \vert
\gamma \vert} \right)^{\frac {1}{2}} \left( ue^{i\omega_Ft} -
u^*e^{-i\omega_Ft}\right)
\end{equation}

$$\tau \,\, = \,\, \vert \delta \omega \vert t$$

\noindent
The equation of motion in terms of the variable {\it u} following from
(3) - (5) takes the form

\begin{equation}
\frac {du}{d\tau} \,\, = \,\, \upsilon \,\, + \,\, \eta \tilde f (\tau)
\end{equation}

$$\upsilon \,\, \equiv \,\, \upsilon (u,u^*) \,\, = \,\, - \eta
u \,\, + \,\, iu(\vert u \vert^2 - 1) \,\, - \,\, i\beta^{\frac {1}{2}}$$

\noindent
where

\begin{equation}
\eta \,\, = \,\, \Gamma /\delta \omega, \qquad \beta \,\, = \,\,
\frac {3 \gamma F^{2}}{32\omega^3_F \delta \omega^3}
\end{equation}

\noindent
are respectively the reduced damping coefficient and the dimensionless
field intensity.  The equations (7) as written corresponds to the particular
case

\begin{equation}
\delta \omega > 0, \qquad \gamma > 0
\end{equation}

\noindent
The generalization to the case where the signs of $\delta \omega$ and $\gamma$
are arbitrary is straightforward.  We note that bistability can occur
only for $\gamma \delta \omega > 0$; simultaneous change in the signs of
$\gamma$ and $\delta \omega$ will result in mirror reflection of the spectra
considered below with respect to $\omega_F$.

The random force $\tilde f (\tau)$ appearing in (6) is proportional to
$f(t)$ in (3),

\begin{equation}
\tilde f (\tau) \,\, = \,\, - i \left( \frac {3 \gamma}{8 \omega^3_F
\Gamma^2 \vert \delta \omega \vert} \right)^{\frac {1}{2}} {\rm exp}
(-i \omega_F t)  f(t)
\end{equation}

\noindent
and, if {\it f(t)} is Gaussian white noise of characteristic intensity
{\it B}, such that

\begin{equation}
\langle f(t) \,\, f(t^{\prime}) \rangle \,\, = \,\, 2 \Gamma B \delta
(t - t^{\prime})
\end{equation}

\noindent
then $\tilde f (\tau)$ is asymptotically a two-component white noise,

$$\langle \tilde f (\tau) \, \tilde f (\tau^{\prime}) \rangle \,\, = \,\,
\langle \tilde f^*(\tau) \tilde f^* (\tau^{\prime}) \rangle \,\, = \,\, 0 \,\,
$$

\begin{equation}
\langle \tilde f (\tau) \tilde f^*(\tau^{\prime}) \rangle = 4 \alpha \delta
(\tau - \tau^{\prime})
\end{equation}

$$\alpha \,\, = \,\, \frac {3 \gamma B}{16\omega^3_F \Gamma}$$

\noindent
The correlator $\langle {\rm Re} \tilde f (\tau) {\rm Im} \tilde f
(\tau^{\prime}) \rangle$ is fast-oscillating; the slow variables $u, u^*$
therefore perceive the components Re $\tilde f (\tau)$, Im $\tilde f(\tau)$
as independent white noises of equal intensity; which is why the correlators
$\langle \tilde f (\tau) \tilde f (\tau^{\prime})\rangle$ and
$\langle \tilde f^*(\tau) \tilde f^*(\tau^{\prime})\rangle$ are
set equal to zero in (11).  We also note that the relations (11) can be
asymptotically fulfilled even where the initial noise {\it f(t)} is not
$\delta$-correlated; it suffices that its correlation time be small as
compared with the \lq\lq slow" process times $\vert \delta \omega \vert^{-1},
\Gamma^{-1}$ (but not necessarily as compared with $\omega^{-1}_F$ [3]).

The dynamics of the system (6) depends on the values of the three dimensionless
parameters involved: $\eta$, $\beta$ and $\alpha$.  We shall assume the
noise to be weak, so that

\begin{equation}
\alpha \ll 1
\end{equation}

\noindent
To zeroth order in $\tilde f(\tau)$, Eq (6) describes the autonomous motion
(note that we consider it in the frame rotating with the frequency of the
external field) of a system with one degree of freedom and, correspondingly,
with two dynamical variables {\it u} and $u^*$ (or Re {\it u} and Im {\it u}).
The stationary solutions of the equation $du/d\tau = \upsilon$ give, in
accordance with (5), the states of steady forced vibration of the oscillator.
The values of the complex envelope {\it u} in the steady states follow from
the relation $\upsilon$ = 0, and are given by

$$u_j = \sqrt {\beta} (\vert u_j \vert^2 \,\, - \,\, 1 \,\, + \,\,
i\eta)^{-1}, \quad \phi (\vert u_j\vert^2) = 0$$

\begin{equation}
\phi (x) \,\, = \,\, x(x - 1)^2 \,\, + \,\, \eta^{2} x - \beta
\end{equation}

\noindent
where $j$ enumerates the real roots of the cubic equation (13) and can
take on the values 1, 2 or 3.  Eq (13) is readily interpreted.  As a result of
nonlinearity, the frequency of the eigenvibrations of the oscillator depends
on their amplitude {\it a} as

$$\omega(a) \,\, \simeq \,\, \omega_0 \,\, + \,\, \frac {3}{8} \gamma a^2
/\omega_0$$

\noindent
On substitution of this expression into the well-known expression for the
amplitude of the forced vibrations of a linear damped oscillator of
frequency $\omega (a)$

$$a^2 \,\, = \,\, \frac {F^2}{[\omega^2_F \,\, - \,\, \omega^2(a)]^2 \,\, +
\,\, 4\omega^2_F \Gamma^2}$$

\noindent
with account taken of the relation $a^2 \,\, = \,\, \frac {8}{3} \vert
u \vert^2 \omega_F \vert \delta \omega \vert /\vert \gamma \vert$, which
follows
immediately from (5), one simply obtains the cubic equation

$$\phi(\vert u \vert^2) \,\, = \,\, 0$$

\noindent
In the parameter range where this equation has three real roots, the
oscillator is bistable: the forced vibrations with the smallest ({\it j} = 1
in (3)) and the largest ({\it j} = 2) amplitudes are stable; there is
also the unstable steady state, {\it j} = 3, with an intermediate value
of $\vert u_j \vert^2 \propto a^2_j$.  The phase of the stable forced
vibrations is determined by the argument of $u_j$ in (13).  The range of
$\beta, \eta$ for which (13) has three solutions, and thus bistability
occurs, corresponds to the approximately triangular region bounded by the
full curves $\beta^{(1, 2)}_B (\eta)$ of Fig 1, i.e. the bifurcation curves.

Thus, as the amplitude of the periodic force is gradually increased from a
small value at fixed frequency (see e.g. vertical line a-a$^{\prime}$ in Fig
1), the system moves from monostability (one small limit cycle), to bistability
(two possible limit cycles of different amplitude), and then back again to
monostability (one large limit cycle).  Some analogy can be drawn between the
bistability and the liquid/gas coexistence region of a Van der Waals system.
As the spinode point, which corresponds to the Van der Waals critical point,
is approached, the two stable (and one unstable) solutions of (13) coalesce
and the amplitudes of the large and small limit cycles (liquid
and gaseous phases) correspondingly approach each other, to become
indistinguishable at the spinode (critical point) itself.  Consequently,
just as in the Van der Waals case, it is possible to move quasistatically
from an initial state that is a small limit cycle (gaseous state) to a final
state that is a large limit cycle (liquid state) without undergoing the
analogue of a first order phase transition or passing through a mixed-phase
coexistence region: all that is necessary is to take a route through the
parameter space that passes outside the spinode (the critical point).

The above analysis makes sense provided that the basins of attraction
are smooth and regular throughout the region of phase space likely
to be visited by the system.  The basins have been computed within the
bistable regime (see Fig 1) in the absence of noise, and are shown as
Poincar\'e sections (values of $\dot {q}, q$ for $t = 2 \pi n \omega^{-1}_F +
\phi_0$)
in Fig 2. We emphasize that the data in Fig 2 refer to the initial
oscillator described by (3) with $f (t)$ = 0.  In addition to the
dimensionless parameters $\eta$, $\beta$, this system is characterised
by the parameter $\Gamma / \omega_F$ which, in the present case, was
set as $\Gamma / \omega_F$ = 0.0184.  The results were obtained
by the usual \lq\lq grid of starts" method [15], allowing the system
to evolve from different starting points in the ($\dot q, q$) phase space and
noting in each case the attractor to which it was drawn.  Thus, all
starts in white areas lead to the large amplitude attractor (the $\bullet$
in the white area), and all starts in black areas lead to the small
amplitude attractor (the $\bullet$ in the black area).  Fig 2 shows the
evolution of the basins with increasing $\beta$ for fixed $\eta$ as Poincar\'e
sections for the same phase.  It is intuitively reasonable that the black
basin (for the small amplitude attractor) should be dominant at small $\beta$,
just within the region of bistability.  As $\beta$ increases, the white basin
(for the large amplitude attractor) grows until the central
regions of the two basins have become equal in area.  With
further increase of $\beta$ the black basin continues to shrink, finally
disappearing at the upper boundary (Fig 1) of the bistable region.

The most important feature of Fig 2 for present purposes is that it confirms
the basins to be (within the range and resolution of the computations)
simple, smooth, and regular as already stated above.  The shapes and
positions of the attractors are close to those given by the approximate
equations of motion (6) in the absence of noise; we note that the latter
equations do not display chaos or fractal boundaries.  Because we are
interested in the regime of weak noise intensity for which the system only
makes {\it occasional} transitions between the attractors, we can be confident
that it spends almost all of its time in the close vicinity of either one or
other of them, and that the probability of fluctuations carrying it out to
regions of phase space where the basins might be irregular or fractal (far
beyond the range plotted in Fig 2) is exponentially small.

Finally in this section, we draw attention to the importance of a slightly
different model, closely  related to (3), that is likely to be more
easily realised in experiments on systems excited by laser radiation with
a randomly varying amplitude:

\begin{equation}
\ddot q + 2 \Gamma \dot q + \omega^2_0 q + \gamma q^3 = [F + f(t)]
\cos \omega_F t
\end{equation}

\noindent
The significance of (14) arises because of the very high value of the
driving frequencies $\omega_F$ in optical experiments, which means
that external noise $f (t)$ introduced from a conventional noise generator
will in practice be far from white; indeed the cut-off frequency of the
generator is likely to be much {\it smaller} that $\omega_F$.

Nevertheless, transforming (14) to the rotating frame again gives
Eq (6), except that the noise is now given by

$$\tilde  f (\tau) = -\frac{i}{2} \left( \frac{3\gamma}{8 \omega^3_F \Gamma^2
\vert \delta \omega \vert}\right)^{\frac {1}{2}} f (t)$$

\noindent
plus a term which varies as $f (t) \exp (2i\omega t)$. The limited
spectral width of $f(t)$ that we have assumed implies that this term will
only have a very small effect on the equation of motion of the slow
variable $u(t)$.  It is the high-frequency components of the noise that
determine the random dynamics of a nearly-resonantly-driven underdamped
nonlinear oscillator; the components of the noise with frequencies far from
$\omega_0$ are filtered out.  In contrast to $\tilde f (\tau)$ (9), the new
$\tilde f (\tau)$ has correlator

$$\langle \tilde f (\tau) \tilde f (\tau^{\prime}) \rangle =
\langle \tilde f^* (\tau) \tilde f^* (\tau^{\prime}) \rangle =
- \alpha \delta (\tau - \tau^{\prime})$$

$$\langle \tilde f (\tau) \tilde f^* (\tau^{\prime}) \rangle =
\alpha \delta (\tau - \tau^{\prime})$$

\noindent
i.e. instead of two independent components, the new $\tilde f (\tau)$ has
only one.  Nonetheless, the analysis presented below can be easily extended
to give similar results for the system (14).  Modulation of the
periodic driving force by noise has pushed the effect of low frequency noise
into the high frequency range.

\vskip 0.3truecm
\noindent
\underbar {B.  Transition probabilities and the spectral density of
fluctuations}

The most obvious effects of noise on the behaviour of the oscillator are,
first, the onset of fluctuations about the stable states and, secondly, the
occurrence of fluctuation-induced transitions between the states.  Provided
that the noise is weak, in accordance with (12), the system will spend most of
its time in the close vicinity of one of the stable states: only very rarely
will a sufficiently large fluctuation occur to cause a transition to the other
stable state.  The dependences of the probabilities $W_{ij}$ of the
transitions on the characteristic noise intensity are of the activation type.

\begin{equation}
W_{ij} = {\rm const.} \times {\rm exp} (-R_i/\alpha)
\end{equation}

\noindent
The activation energy $R_i$ for the transition from state {\it i} is given
by the solution of a variational problem: the corresponding variational
equations and the algorithm for their numerical solution are discussed in the
Appendix.  The resultant dependences of $R_i$ on $\beta$ for the lower
({\it i} = 1) and higher ({\it i} = 2) amplitudes of the forced vibrations
in the limit of small reduced damping $\eta$ were considered in [3].  Numerical
results for four values of $\eta$ are shown by the circles in Fig
3(a)-(d).  It is evident that $R_1$ decreases, and $R_2$ increases,
monotonically with increasing $\beta$ {\it i.e.} with the characteristic
resonant field intensity.  For the values of $\beta$ corresponding to the
upper and lower bifurcation lines $\beta_B^{(1,2)}$ in Fig 1, $R_1$ and $R_2$
respectively vanish (as the states 1 and 2 coalesce with the saddle
point and then disappear).  The dependence of $R_i$ on $\beta$ for $\beta$
close to $\beta_B^{(i)} (\Omega)$ is universal, $R_i = G_i (\eta) \vert
\beta - \beta_B^{(i)} (\eta) \vert^{3/2}$, and is shown by the full lines.
(The explicit form of $G_i$ has been considered earlier, cf [3]).  The
numerical and asymptotic results are in good agreement for not too small
$\eta$, where the optimal path of the escape (in the rotating frame (5):
see Appendix) is not a small-step spiral.  For small $\eta$, however,
the numerical algorithm is not accurate enough and results in the
discrepancies seen in Fig 3(a); as discussed in the Appendix, the data in
this range can better be obtained in a different way.  The dependence of
$R_{1, 2}$ on the frequency detuning $\eta$ for $\beta$ lying in the central
part of the interval $(\beta_B^{(1)} (\eta), \beta_B^{(2)} (\eta)$) is rather
sharp, especially at small $\eta$ where [3] $R_{1,2} \propto \eta$.  As $\eta$
approaches its critical value (the spinode point in Fig 1) given by

$$\eta_K^{-1} \,\, = \,\, \sqrt {3}, \qquad \beta_K \,\, = \,\, 8/27$$

\noindent
the values of $R_{1,2}$ decrease rapidly [3], as ($\eta - \eta_K)^{-2}$.
Here, too, the numerical and analytic results are in good agreement.  It
is evident that, as $\eta$ approaches $\eta_K$, the range of $\beta$
over which $R_i$ is well-described by the asymptotic law
$\vert \beta - \beta_B^{(i)} (\eta) \vert^{3/2}$ increases relative to the
total range of bistability $\vert \beta_B^{(2)} (\eta) - \beta_B^{(1)}
(\eta)\vert$.  Qualitatively, this is related to the fact that the optimal
path of the escape is approaching a straight line on the plane $(u^{\prime},
u^{\prime \prime})$, and it becomes nearly straight for all $\beta, \eta$
close to the spinode point.

The analytic and calculated values of $R_1$ and $R_2$ allow us to plot on
Fig 1 the dashed curve $\beta_0(\eta)$ specifying the points at which the
activation energies are equal, closely approximating the line of the kinetic
phase transition (KPT) at which the populations are equal,

\begin{equation}
R_1(\beta, \eta) \,\, = \,\, R_2 (\beta, \eta), \quad \beta = \beta_0 (\eta)
\end{equation}

\noindent
(Note that the criteria $R_1 = R_2$ differs from $w_1 = w_2$ only by virtue
of variations in the prefactor $\sim \alpha$ in (15); the criteria become
identical as $\alpha \to 0$).  For parameter values far from this curve,
the transition probabilities $W_{12}$ and $W_{21}$ are seen from (15) to
differ exponentially strongly.  Correspondingly, the stationary populations
$w_1, w_2$ of the states as given by (2) are also exponentially different,

$$w_1/w_2 \,\, \propto \,\, {\rm exp} [(R_1 - R_2)/\alpha]$$

\noindent
Only in the close vicinity of the $\beta_0(\eta)$ curve will the transition
probabilities and stationary populations be comparable and it is here,
therefore, that one may expect to observe the characteristic steady state
fluctuation phenomena associated with transitions between the attractors;
we do not consider here the transient fluctuation effects that arise when,
for example, the parameters are swept through the bifurcation lines in
Fig 1.

A revealing characteristic property of a fluctuating system is its spectral
density of fluctuations (SDF).  The SDF of the co-ordinate of a
periodically driven oscillator, $Q(\omega)$, is given by

$$Q(\omega) \,\, = \,\, \frac {1}{\pi} \, {\rm Re}\int^{\infty}_{0} dt \,
{\rm exp} (i \omega t) \tilde Q (t)$$

\begin{equation}
\tilde Q (t) \,\, = \lim_{T \to \infty} \,\, \frac {1}{2T}
\int^T_{-T} d \tau [q(t + \tau) - \langle q(t + \tau) \rangle] [q(\tau) -
\langle q (\tau) \rangle ]
\end{equation}

\noindent
We note that a periodically driven system is in general nonergodic, so that
$\tilde Q (t)$ is not equal to the time correlation function

$$\langle [q(t + \tau) - \langle q(t + \tau) \rangle][q(\tau) - \langle q
(\tau) \rangle] \rangle$$

\noindent
defined in terms of ensemble averaging $\langle .... \rangle$; in fact, the
latter quantity oscillates with $\tau$ at frequency $\omega_F$, as can be
seen from (5), (6); $\tilde Q (t)$ actually corresponds to this quantity
smoothed over $\tau$.

The ensemble-averaged value of the coordinate, $\langle q (t) \rangle$, is
equal to the value of $q(t)$ averaged over equal instants of time
modulo $2\pi /\omega_F$,

$$\langle q (t) \rangle = \lim_{N \to \infty} N^{-1} \sum\limits^{N-1}_{n=0}
q (t + 2 \pi n \omega^{-1}_F)$$

In the case of weak noise, two principal contributions to $Q(\omega)$ can
be identified [12, 18].  The first of these arises from
small fluctuations about the stable states.  It is equal to the sum over the
states {\it j} of the corresponding partial SDFs, $Q_j(\omega)$, weighted
by the state populations $w_j$ given by (2) (cf also [19]).  The second
contribution $Q_{tr}(\omega)$ comes from the (relatively infrequent)
fluctuational transitions between the states.  Thus (cf [5])

\begin{equation}
Q(\omega) \,\, = \,\, \sum \limits_{j}^{} w_j Q_j (\omega) \,\, + \,\,
Q_{tr} (\omega)
\end{equation}

To calculate the partial SDF, $Q_j(\omega)$, for the state {\it j} when the
noise intensity $\alpha$ is small, it suffices: to linearize the $\upsilon$
term on the right hand side of (6) with respect to small deviations in
$(u -u_j), (u^* - u^*_j)$; to substitute the solution of the resultant linear
equations into (17), taking due account of (5); and to perform averaging.
The result is of the form

$$Q_j(\Omega) \,\, = \,\, \frac {4\omega_F\alpha \Gamma^2}{3 \vert \gamma \vert
\pi} \,\,\frac {(\omega - \omega_F)^2  +  2 (\omega - \omega_F) \Gamma
\eta^{-1}
(2 \vert u_j \vert^2 - 1) + \Gamma^2(\nu^2_j + 2 \eta^{-2} \vert u_j \vert^4)}
{[(\omega - \omega_F)^2 - \Gamma^2 \nu^2_j]^2 \,\, + \,\, 4\Gamma^2 (\omega
- \omega_F)^2}$$

\begin{equation}
\nu^2_j \,\, = \,\, 1 \,\, + \,\, \eta^{-2} (3\vert u_j \vert^2 - 1)(\vert
u_j \vert^2 - 1)
\end{equation}

\noindent
where $\vert u_j \vert^2$ for {\it j} = 1, 2 is given by Eq (13).  It is
evident from (19) that $Q_j(\omega)$ is peaked near the frequency
$\omega_F$ of the external field.  Its intensity will be proportional to the
noise intensity.  The shape of $Q_j(\omega)$ will be discussed in Sec IV.

The second term in (18), $Q_{tr}(\omega)$, can readily be calculated if one
notes that the populations $w_j$ of the stable states fluctuate in time
with a characteristic relaxation time $\left( W_{12} \,\, + \,\, W_{21}\right)
^{-1}$, so that

$$\frac {dw_1(t)}{dt} \,\, = \,\, - \left( W_{12} + W_{21} \right) w_1 (t)
\,\, + \,\, W_{21}$$

\begin{equation}
w_2(t) \,\, = \,\, 1 - w_1(t)
\end{equation}

\noindent
(The values of $w_j$ appearing in (2), (18) correspond to the stationary
solutions of (20)).  In the case of weak noise, these fluctuations can be
shown to result in a contribution to $Q(\omega)$ of

\begin{equation}
Q_{tr}(\omega) \,\, = \,\, \frac {2 \omega_F \vert\omega_F -
\omega_0\vert}{3\pi
\vert \gamma \vert}  \vert \langle u_1 \rangle - \langle u_2 \rangle \vert^2
w_1 w_2 \frac {W_{12} + W_{21}}{\left( W_{12} +  W_{21} \right)^2  +
(\omega - \omega_F)^2}
\end{equation}

\noindent
Here $\langle u \rangle_j$ denotes the ensemble average value of $u$ in the
state $j$. In the zero noise limit $\langle u \rangle_j$ is simply $u_j$.  For
the purposes of comparison with experiments performed at finite
noise intensity, $\langle u \rangle_j$ can be expanded as a perturbation
series in the small parameter $\alpha$.  To first order we obtain

$$\langle u \rangle_j = u_j + \langle \delta u_j \rangle$$

$$\langle \delta u_j \rangle = \frac {2\alpha u_j}{\eta^3 u^4_j}
\left\{ i \eta [ 2 \eta^2 + 3 \vert u_j \vert^4 - 6 \vert u_j \vert^2
+ 2] - (3 \vert u_j \vert^2 - 2)(\eta^2 + 2 \vert u_j \vert^4 -
3 \vert u_j \vert^2 + 1)\right\}$$

\noindent
We note that the spectral peak $Q_{tr}(\omega)$ is extremely narrow:  its
width is determined by the transition probabilities, so that it is
exponentially small and much smaller than the damping parameter $\Gamma$ which
determines the \lq\lq dynamical" relaxation of the oscillator towards either
of its stable states.  The product $w_1 w_2$, which determines the intensity
of $Q_{tr}(\omega)$, can be seen from (2), (15) to be exponentially small for
almost all values of $\beta, \eta$, with the exception of those within
the very narrow range (the phase transition region) where $w_1 \sim w_2 \sim
1$.  Thus, the onset of the fluctuational transition-induced spectral
peak $Q_{tr}(\omega)$ is a specific phase transition phenomenon (see Sec IV).

\vskip 0.3truecm
\noindent
\underbar {C.  The susceptibility and high frequency stochastic resonance}

The effect of a weak trial periodic force on thermal equilibrium systems,
is the onset of vibrations at the frequency of the force; their amplitude is
characterised by a susceptibility, which can be expressed in terms of the
SDF via the fluctuation dissipation theorem [20]. If the system is being
driven by a strong periodic force $F {\rm cos} \omega_F t$, so that it is
far from thermal equilibrium, the additional weak force
$A \exp (-i \Omega t)$ gives rise to vibrations not only at its own
frequency $\Omega$, but also at combination frequencies $\vert \omega \pm
\omega_F \vert, \vert \Omega \pm 2 \omega_F \vert$, ...

We shall consider the linear response of the bistable oscillator to a nearly
resonant trial force with a frequency $\Omega$ close to $\omega_0, \omega_F$:

$$\vert \Omega - \omega_0 \vert, \quad \vert \Omega - \omega_F \vert \,\,
\ll \omega_F$$

\noindent
In this case, a strong response is to be expected. It will be most pronounced
at the frequency $\Omega$ and at the nearest resonant combination frequency,
which is $2\omega_F- \Omega$.  Thus, one can seek the trial force-induced
modification of the ensemble-averaged coordinate {\it q} in the form

\begin{equation}
\delta \langle q(t) \rangle \,\, \simeq \,\, \chi (\Omega) A \exp
(-i \Omega t) \,\, + \,\, X(\Omega) A {\rm exp} [i(2\omega_F - \Omega) t]
\end{equation}

\noindent
That is, we may suppose that the linear response is characterised by two
coefficients (generalised susceptibilities), $\chi(\Omega)$ and $X(\Omega)$.
The absorption/amplification of the trial field is characterised by Im
$\chi (\Omega)$.  It was shown in [3] that, in the vicinity of the KPT,
interesting features occur in the absorption spectrum.

To calculate the susceptibilities we transform to the slow variables $u, u^*$
(5) in the equation of motion (3) with the additional force $A \exp (-i
\Omega t)$ added to the right hand side.  The resultant equations for
$u, u^*$ take the form

$$\frac {du}{d\tau} \,\, = \,\, \upsilon(u, u^*) \,\, + \,\, \eta
\tilde f (\tau)$$

\begin{equation}
\frac {du^*}{d \tau} \,\, = \,\, \upsilon^*(u, u^*) \,\, + \,\, \eta
\tilde f^* (\tau) \,\, + \,\, i \tilde {A} (\tau)
\end{equation}

$$\tilde {A} (\tau) \,\, = \,\, \left[ \frac {3 \gamma}{8 \omega^3_F
\vert \delta \omega \vert^3} \right]^{\frac {1}{2}} \, A \exp \left[
\frac {-i(\Omega - \omega_F) \tau}{\vert \delta \Omega \vert}\right]$$

\noindent
It is evident from (23) how the second term in (22) arises: it is due to the
addition $\propto {\rm exp} [-i(\Omega - \omega_F)t]$ to {\it u}, which is then
multiplied by ${\rm exp} (i \omega_F t)$ when {\it q(t)} is calculated in (5).

If the random force $\tilde f(\tau)$ is weak, the main effects of the
additional term $\propto \tilde {A} (\tau)$ in (23) are: (i) to cause
small amplitude periodic vibrations of $u, u^*$ about their stable values
$u_j, u^*_j$; and (ii) via the change in the probabilities of fluctuational
transitions, to modulate periodically the populations of the stable states.
These effects give rise to expressions for the generalised susceptibilities
of the form

$$\chi (\Omega) \,\, = \,\, \sum \limits_{j}^{} w_j \chi_j(\Omega) \,\,
+ \,\, \chi_{tr} (\Omega)$$

\begin{equation}
X(\Omega) = \sum\limits_{j}^{} w_j X_j (\Omega) \,\,
+ \,\, X_{tr} (\Omega)
\end{equation}

\noindent
where $\chi_j$, $X_j$ are the partial susceptibilities related to the
corresponding vibrations about the stable states, and $\chi_{tr}$, $X_{tr}$
are related to the trial force-induced redistribution over the states.

The partial susceptibilities can readily be calculated by linearising (23)
near the stable states, yielding

$$\chi_j (\Omega) \,\, = \,\, \frac {i}{2\omega_F} \quad \frac {\Gamma - i
(\Omega - \omega_F) - i(2 \vert u_j \vert^2 - 1) (\omega_F - \omega_0)}
{\Gamma^2\nu^2_j - 2i\Gamma (\Omega - \omega_F) - (\Omega - \omega_F)^2}$$

\begin{equation}
X_j(\Omega) \,\, = \,\, \frac {-1}{2\omega_F} \quad \frac {u^2_j
(\omega_F - \omega_0)}{\Gamma^2 \nu^2_j - 2i\Gamma (\Omega - \omega_F) -
(\Omega - \omega_F)^2}
\end{equation}

The effective modulation of the transition probabilities by the trial field
$A$, which gives rise to the second term on the right hand side of
each of Eq (24), arises when its frequency $\Omega$ is very close to
$\omega_F$, so that

$$\vert \Omega - \omega_F \vert \,\, \ll \,\, \Gamma, \,\,\, \vert
\omega_F - \omega_0 \vert$$

\noindent
In this case, the trial field smoothly raises and lowers the effective
\lq\lq barrier" between the stable states with the period $2\pi / \vert \Omega
- \omega_F \vert$, so that the activation energies $R_1, R_2$ of the
fluctuational transitions vary periodically in time [3].  The corresponding
additions to $R_j$ are given in the Appendix.  In turn, they give rise to
periodic additions to the transition probabilities $W_{ij}$ (15) and hence to
the population $w_j$ of the stable states (20).  The final expression for the
redistribution-induced additions to the generalised susceptibilities is

$$\chi_{tr} (\Omega) \,\, = \,\, \frac {w_1w_2}{2 \omega_F(\omega_F -
\omega_0)} (\langle u \rangle^*_1 - \langle u \rangle^*_2)(\frac {\mu_1 -
\mu_2}{\alpha}) \left[ 1 - \frac {i(\Omega - \omega_F)}{W_{12} +
W_{21}}\right]^{-1}$$

\begin{equation}
X_{tr} (\Omega) \,\, = \,\, \frac {\langle u \rangle_1 - \langle u
\rangle_2}{\langle u \rangle_1^* - \langle u\rangle^*_2} \,\,
\chi_{tr} (\Omega)
\end{equation}

$$\mu_j \,\, = \,\, \sqrt {\beta} \,\, \left( \frac {\partial R_j}{\partial
\beta}\right)$$

\noindent
It is evident from (26) that the susceptibilities $\chi_{tr} (\Omega),
X_{tr} (\Omega$) are large only within the range of parameters $\beta, \eta$,
close to the kinetic phase transition, where the populations $w_1, w_2$ of
the stable states are of the same order of magnitude.  The characteristic
range of the frequency $\Omega$ of the trial field within which these
susceptibilities are large is determined by the transition probabilities.
Consequently, it increases exponentially with increasing noise intensity (cf
[12]).  This property gives rise to {\it stochastic resonance} [8] i.e.
to an increase of the signal/noise ratio (SNR) with increasing noise [21]
which, as shown below, occurs in the present system for a high frequency signal
$\Omega \simeq \omega_F \gg \Gamma$.  To calculate the ratio,
we note from (22) that the signal induced by a real field $A {\rm cos} \Omega
t$ is given by

\begin{equation}
\delta \langle q(t) \rangle \,\, = \,\, A \, {\rm Re} \left\{ \chi (\Omega)
{\rm exp} (-i \Omega t) \,\, + \,\, X (\Omega) {\rm exp} \left[ i
(2\omega_F - \Omega) t \right] \right\}
\end{equation}

\noindent
Such a signal corresponds to the appearance of $\delta$-shaped spikes in the
power spectrum of the oscillator at frequencies $\Omega$ and (2$\omega_F -
\Omega$).  This can be seen from (17) if (27) is added to $q(t)$, $q(t + \tau)$
but not to $\langle q (t) \rangle$.  (The latter quantity is included in (17)
to subtract the $\delta$-function in $Q(\Omega)$ at the frequency $\omega_F$
of the strong field).  It is evident from (16), (17) that the ratios {\it P}
and ${\cal P}$ of the strengths (areas) of the spikes at frequency $\Omega$
of the trial field, and at the combined frequency (2$\omega_F - \Omega$), to
the power spectrum in the absence of noise are given by

\begin{equation}
P = \frac {S}{Q(\Omega)}, \qquad {\cal P} = \frac {S}{Q(2\omega_F - \Omega)}
\end{equation}

$$S = \frac {1}{4} A^2 \vert \chi (\Omega) \vert^2, \qquad S = \frac {1}{4}
A^2 \vert X (\Omega) \vert^2$$

It follows from (25) that the \lq\lq partial" susceptibilities
$\chi_j(\Omega$),
$X_j(\Omega)$ are independent of noise for weak noise, whereas the partial
contributions to the SDF $Q_j(\Omega)$ increase linearly with the noise
intensity.  Far from the phase transition region, therefore, where the
fluctuational transition contributions to the susceptibilities and SDF are
small, the quantities {\it P} and ${\cal P}$ decrease with increasing noise.
Within the phase transition range, on the other hand, for small $\vert \Omega -
\omega_F \vert \sim W_{ij}$, the main contribution to $\chi(\Omega)$,
$X(\Omega)$ and $Q(\Omega)$ comes just from the transitions (26), (21).  This
is because their ratio to the corresponding partial contributions is inversely
proportional to the (small) noise intensity parameter $\alpha$.  If only
$\chi_{tr}$, $X_{tr}$ and $Q_{tr}$ are taken into account in (28), one
obtains the corresponding quantities $P_{tr}$, ${\cal P}_{tr}$

\begin{equation}
P_{tr} \,\, = \,\, {\cal P}_{tr} \,\, = \,\, A^2 \, \frac {3\pi \vert \gamma
\vert}
{32\omega^3_F \vert \omega_F - \omega_0\vert^3} \left( \frac {\mu_1 -
\mu_2}{\alpha}
\right)^2 \, \frac {W_{12}W_{21}}{W_{12} + W_{21}}
\end{equation}

According to (29), the quantities $P_{tr}, {\cal P}_{tr}$ are independent of
frequency.  At the same time, they can be seen from (15) to  increase
exponentially with increasing noise intensity.  This implies the onset of high
frequency stochastic resonance within the phase transition range, not only
at the frequency of the trial field, but also at the combinational frequency
(2$\omega_F - \Omega$).  In fact, the ratio ${\cal P}$ is rather different
from the quantity usually considered in the context of SR, because no
force is being applied at the frequency $\vert \Omega - 2 \omega_F \vert$:
the signal is induced by mixing, in a nonlinear system, of the forces at
frequencies $\Omega$ and $\omega_F$.  In relation to nonlinear optics
[22], the phenomenon can be regarded as a type of highly selective,
resonant, four-wave mixing (actually, multiwave mixing because the
effect is not just proportional to the squared amplitude $F$ of the
strong field).

We would emphasize that stochastic resonance occurs only within the phase
transition region.  When the parameters $\beta, \eta$ of the oscillator
are far from this region, the contributions $\chi_{tr}$, $X_{tr}, Q_{tr}$ to
the susceptibilities and the SDF in the absence of the trial force are
exponentially small: {\it P} and ${\cal P}$ differ
markedly from $P_{tr}, {\cal P}_{tr}$, therefore, and decrease with increasing
$\alpha$.  The dependences of {\it P} and ${\cal P}$ on $\alpha$, as given by
(18), (19), (21), (24)-(26) and (28), will be compared with the results
of the analogue electronic experiments in Section IV below.

\vskip 0.3truecm
\centerline {III.  ANALOGUE ELECTRONIC EXPERIMENTS ON}
\centerline {THE PERIODICALLY DRIVEN OSCILLATOR}

In order to test the theoretical predictions of the preceding section, and to
find out whether they were applicable to a real physical system described
by the model equation (3), a series of analogue electronic experiments was
undertaken.  The basis of the analogue technique has been described in detail
elsewhere [23], together with a discussion of its advantages and disadvantages.
In essence, it is extremely simple.  An electronic model of the stochastic
differential equation under study is built using standard analogue components
(operational amplifiers, multipliers etc).  This is then driven by
stochastic and periodic forces, as appropriate, and its response is
analysed with the aid of a digital data processor.

The circuit used to model (3) is shown in (slightly simplified) block
form in Fig 4.  It was designed and scaled in the standard [23] way so as
to optimise use of the dynamic range of the active components.  Thus, the
actual equation simulated (see Fig 4) was the integral form of

$$R_1 C_1 R_4 C_2 \ddot x + \frac {R_1}{R_3} R_4 C_2 \dot x = -
\frac{R_1}{R_5} x - \frac {R_1}{R_6} \frac {x^3}{20}$$

$$+ \frac {R_1}{R_2} F^{\prime} \sin \omega^{\prime}_F t +
\frac{R_1}{R_0} A^{\prime} \sin \Omega^{\prime}t + \eta (t)$$

\noindent
with

$$R_1 = R_4 = R_5 = 2R_6 = \frac{R_2}{10} = \frac{R_0}{10} = 100 k\Omega$$
$$R_3 = 2.5 {\rm M}\Omega$$
$$C_1 = C_2 = 1 {\rm nF}$$
$$\tau = R_1 C_1 = R_4 C_2$$
$$2\Gamma = R_1/R_3$$

\noindent
Thus the equation actually simulated was

$$\tau^2 \ddot x + 2\Gamma \tau \dot x = -x - \frac{x^3}{10} +
\frac{F^{\prime}}{10} \sin \omega_F^{\prime} t + \frac {A}{10}
\sin \Omega^{\prime} t + \eta (t)$$

\noindent
which, after the scaling

$$t \to t^{\prime}/\tau, \quad \omega_F \to \omega^{\prime}_F \tau, \quad
\Omega \to \Omega^{\prime} \tau, \quad F \to F^{\prime}/10, \quad
A \to A^{\prime}/10$$

\noindent
goes over into (3) with $\omega_0 = 1$, $\alpha$ = 0.1.  Provision was made
for measuring either the coordinate $q(t)$ or the energy
$E = \frac {1}{2} \dot q^2 + \frac {1}{2} q^2 + \frac {1}{4} \gamma q^4$.

The circuit model was driven with a sinusoidal periodic force from a
Hewlett-Packard 3325B frequency synthesizer.  Its response, a time-varying
voltage representing $q(t)$, was digitized (12-bit precision) typically
in 1K or 2K blocks and analysed using a Nicolet LAB80 data-processor; for
the experiments on high frequency stochastic resonance, where larger $q(t)$
data blocks were required (see below) a Nicolet 1280 data-processor was
used.

As expected, the model was found to display bistability within a certain
range of forcing amplitude and frequency: its response $q (t)$ in the
absence of noise for a given set of parameters, shown in Fig 5(a) and (b),
could have either of two distinct amplitudes, corresponding to the
two coexisting periodic attractors discussed in Section II.  An
inherent experimental difficulty of the measurements lay in the
accurate determination of $\beta$, on account of the $\vert \omega_F
- \omega_0 \vert^3$ term in the denominator.  Because, for the
region of interest, the forcing frequency $\omega_F$ is very close to
the natural (zero amplitude) frequency $\omega_0$ of the oscillator,
a very small error (typically $\pm$ 1\%) in $\omega_0$ inevitably
results in a much larger error (typically $\pm$ 40\%) in the value of $\beta$.
For this reason, rather than attempting to determine $\omega_0$ from
the nominal component values or, directly, by a resonance experiment,
its value was established precisely by measurement of the range of
bistability at a single value of $\eta$.  Once this had been done,
the rest of the region of bistability could be mapped out over the whole
range of $\eta$, resulting in the square data points of Fig 6.   They are
seen to be in satisfactory agreement with the theoretical prediction of
Section II (full curves).

The energy $E = \frac {1}{2} \dot q^2 + \frac {1}{2} q^2 +
\frac {1}{4} \gamma q^4$ of the oscillator (apart from the coupling energy
to the force), measured as a function  of $q$ in the
absence of noise for each of the attractors, is shown in Fig 5(b).
Both $E$ and $q$ are periodic functions of time, so that the energy can in
principle take on several values for any given $q$, depending on how many
times $\dot q$ becomes zero during one period $2\pi / \omega_F$.  For the
Duffing oscillator in the range of parameters considered here, where the
nonlinearity is relatively small (see Section IIA), $\dot q$ was zero twice
within a period and therefore $E$ could take on not more than two values
for a given $q$.  Since, for small enough nonlinearity, the steady
vibrations $q(t)$ have components at the odd overtones only, i.e. at the
frequencies $\omega_F, 3\omega_F, 5\omega_F \dots$, there is an
additional symmetry: $q(t + \pi /\omega_F) = -q(t)$, $E (t + \tau/\omega_F)
= E(t)$.  Consequently, the energy is a unique function of the
coordinate $q$ on an attractor.  We emphasize, however, that $E$ is
not conserved: the oscillator acquires energy from the periodic driving
force and dissipates it through friction.  It is evident from Fig 5(b) that
the curvature of $E(q)$ is relatively small, providing a clear indication
that the amplitudes of the harmonics of $E(t)$ are also small; in
the approximation (6), they have been ignored.  The smearing (thickness)
of the $E(q)$ lines in Fig 5(b) is an experimental artefact: the values of
$E$ and $q$ were recorded at discrete intervals of time, and
neighbouring pairs of values have been connected by straight lines
(cf Fig 7(b) below).

When noise was applied to the driven oscillator, fluctuations about
the attractors and occasional transitions between them were observed.
Fig 7(a) shows an experimental example of one such transition; the
corresponding $E(q)$ plot in Fig 7(b) shows the fluctuations
in energy about each of the attractors, yielding an envelope that
illustrates very clearly the shape of the potential.

Measurement of the transition probabilities between the attractors was not
completely straightforward because there was a small region of overlap
between them in terms of any single variable whether measured, for example,
in terms of $q (t)$  or of $E (q)$.  Thus the determination of sojourn times
[24] on either side of a fixed value of $q$ or $E$ would not have provided the
information sought.  Instead, the mean first passage time (MFPT) was measured
between {\it two} pre-set criterion levels in energy, which were outside the
overlap region and unambiguously within each of the attractors.  Fig 8(a)
illustrates how the apparent MFPT varied when one criterion level was kept
fixed within the lower energy attractor, and the other level was moved through
different values.  There is clearly a plateau region around the noise-free
energy level of the attractor for which the MFPT was independent of level
setting: all of the MFPT measurements to be reported below refer to this
region.

To determine SDFs, a standard fast Fourier transform (FFT) routine was
used to compute the power spectral density of the fluctuations
$q(t) - \langle q (t) \rangle$.  In practice, the ensemble-averaged signal
$\langle q (t) \rangle$ was determined in a preliminary
experiment for each set of parameters, averaging a large number (typically
1000) of blocks of $q (t)$ in order to obtain good statistical quality. This
was possible because of $\langle q (t) \rangle$ being strictly periodic,
with $\langle q (t) \rangle = \langle q (t + 2\pi/\omega_F)\rangle$,
and because the phase of $\langle q (t) \rangle$ was determined with
respect to that of the field $F \cos \omega_F t$.  The resultant
$\langle q (t) \rangle$ was then subtracted from each
subsequent realisation of $q (t)$ before the FFT was applied to find the SDF,
which was
itself ensemble-averaged to produce the final result.

The experiments on high frequency stochastic resonance involved the
application of an additional weak trial force $A \cos (\Omega t)$
to the system, with $\Omega$ very close to the main forcing frequency
$\omega_F$.  In order to resolve the expected (see above) responses
at $\Omega$ and at $\vert 2\omega_F - \Omega \vert$ from the supernarrow
peak at $\omega_F$, it was necessary to use a relatively large block size,
which in practice was set to 8K or 16K using the Nicolet 1280 data processor.

\vskip 0.3truecm
\centerline {IV.  DISCUSSION OF RESULTS}

We now compare the theoretical predictions of Section II with the results
of he analogue experiments described in Section III.
As already noted above, all of the main features expected on the basis
of the theory have been observed in the simulations: for example, the
anticipated bistability of the oscillator was observed and its range
in terms of $\beta, \eta$ was found (see comparison of data points and
theory in Fig 6) to be in agreement with the theoretical predictions;
and with weak noise applied to the system, fluctuational transitions were
observed to be taking place between the stable states.  We now
present a more detailed comparison of experiment and theory considering,
in turn, escape probabilities, spectral densities of the fluctuations and
high frequency stochastic resonance in Subsections A, B and C respectively.

\noindent
\underbar {A.  Transition probabilities}

To characterise the transition probabilities, the average lifetimes
$\langle T_i \rangle$ of the states were measured (with the mean time
$\langle T_i \rangle$ from the initially occupied state {\it i} being
measured in the absence of back-flow as described in Section III, so that

$$\langle T_i \rangle \,\, = \,\, W^{-1}_{ij} \,\, = \,\,- \int^{\infty}_0 t
(dw_i/dt)_0 dt$$

\noindent
with $(dw_i/dt)_0 \,\, = \,\, -W_{ij}w_i$ and $w_i(0) \,\, = \,\, 1$).  Some
typical measurements of the average lifetime $\langle T_i \rangle =
W_{ij}^{-1}$, on a log plot as a function of noise intensity $\alpha$, are
shown in Fig 8(b).  The fact that the data fall on straight lines confirms
that the escape process is of the activation type, as expected on the basis
of Eq (15); the characteristic activation energies $R_i$ of the transitions
for given $\beta, \eta$ can be obtained immediately from the slope in each
case.  Some experimental values of $R_i$, obtained in this way from a
large number of measurements similar to those of Fig 8, are presented in
Fig 9.  In good qualitative agreement with the theoretical predictions (full
curves), the value of $R_1$ for the transition from the lower amplitude
attractor decreases monotonically with increasing dimensionless field intensity
$\beta$, while $R_2$ for the transition from the higher amplitude
attractor correspondingly increases.  Both $R_1$ and $R_2$ increase with the
increase of the frequency detuning parameter $\eta$.  Note that the
experimental errors here are relatively large, due to the problem of measuring
$\eta$, discussed above, and to the effect of small changes $\sim \pm$ 0.5\%
in $\omega_0$ with ambient temperature.

The values of $\beta, \eta$ for which $R_1 = R_2$ lie extremely close to those
values for which $\langle T_1 \rangle \, = \, \langle T_2 \rangle$ [because
the effect of the prefactor in Eq (15) is relatively weak], defining the
kinetic phase transition. The phase transition points obtained from the
experimental data (for $\langle T_1 \rangle = \langle T_2 \rangle$, crosses)
fall close to the position of the theoretical phase transition line (for $R_1
= R_2$, dashed) in Fig 6.  The discrepancy between experiment and theory
becomes somewhat larger near the
spinode point K where the system is very \lq\lq soft" and extremely weak
noise intensities are necessary to make the transition region sufficiently
narrow and the phase transition itself sufficiently sharp.  The influence
of uncertainties in the experimental parameters, and of internal noise in
the active circuit components, becomes even more important here.  Taking due
account of all these factors, it may be concluded that theory and experiment
are in satisfactory agreement.

\vskip 0.3truecm
\noindent
\underbar {B.  Spectral density of fluctuations}

Experimental measurements of the SDF in the vicinity of the oscillator
eigenfrequency (histograms) are presented and compared with theory (full
curves) in Fig 10(a)-(c).  It must be emphasized that the measurements refer
to the spectral density of {\it fluctuations about an ensemble average};
the subtraction of $\langle q (t) \rangle$ from each realisation $q (t)$
prior to computation of $Q (\omega)$ ensures that, when the system remains
on one particular attractor throughout, most of the $\delta$-shaped at
$\omega_F$ (which is the Fourier transform of $\langle q (t) \rangle$)
gets removed.  This is why there is very little sign of a spectral peak at
$\omega_F$ in Figs 10(a) and 10(c).  In the KPT range, however, where jumping
occurs between the attractors, the general appearance of the spectrum is
entirely different.  In fact, the most striking feature of the spectrum is the
supernarrow peak [15] that rises in the phase transition range, where
$\beta \simeq \beta_0 (\eta)$, as seen in Fig 10(b).  Its width is very much
smaller than either of the widths of the other peaks, or the experimentally
determined damping constant $\Gamma$, or the frequency detuning $\omega_F -
\omega_0$ (which are all of the same order of magnitude).  For small noise
intensities $\alpha$, this width is unresolved by the LAB80 data analysis
system {\it i.e.} the peak lies entirely within one \lq\lq bin" of the
data-processor's memory.  It was necessary to increase the noise intensity
substantially in order to spread the peak over two or three bins.

The dependence of the intensity $I$ of the supernarrow peak on the
distance from the phase transition line was found to be exponential, as shown
in Fig 11.  This feature can readily be understood in terms of (2), (15) and
(21).  According to (21), for small noise intensities where $\langle u_i
\rangle \simeq u_i$,

\begin{equation}
I = \int^{\infty}_{-\infty} d \omega Q_{tr} (\omega) \,\,
= \,\, \frac {2 \omega_F \vert \omega_F - \omega_0 \vert}{3 \vert \gamma \vert
} \vert u_1 - u_2
\vert^2 w_1 w_2
\end{equation}

\noindent
Not too far from, but not too close to, the phase transition line where, on the
one hand, $\vert \beta - \beta_0 (\eta) \vert \ll 1$ and, on the
other, the transition probabilities $W_{12}$ and $W_{21}$ differ substantially
from each other, the coefficient $w_1 w_2$ in (21) should behave, according
to (2), (15) as

\begin{equation}
w_1w_2 \,\, \propto \,\, {\rm exp} \left\{ - \vert R_1^{\prime} - R_2^{\prime}
\vert \vert \beta - \beta_0 (\eta) \vert / \alpha \right\}
\end{equation}

\noindent
where $R^{\prime}_i \,\, = \,\, (\partial R_i/\partial \beta)_{\beta = \beta_0
(\eta)}$.  Therefore, the dependence of $I$ on ($\beta - \beta_0$)
should indeed be exponential.  In the immediate vicinity of the phase
transition, where the exponent on the right hand side of (31) is of order
unity, this exponential dependence will be smeared.  Such smearing is clearly
seen in the experimental data (squares, Fig 10), which are in good agreement
with (30)  as indicated by the crosses ($w_1, w_2$ having been taken from
independent measurements of the transition probabilities).  Agreement with (30)
based on $w_1 w_2$ taken from the simple estimate (31), as indicated by the
full lines, is also good: the values of $R_{1, 2}^{\prime}$ in this case were
taken from the slopes of the experimental $R_{1, 2} (\beta)$ measurements
and the prefactor in (31) was taken to be $\frac {1}{4}$ so as to give the
correct maximum value of $w_1 w_2$ at the phase transition point $\beta =
\beta_0(\eta)$ in the limit $\alpha \to 0$.  The cusp-like dependence of the
intensity $I$ of the supernarrow peak in the SDF is a characteristic feature of
the peak, which itself represents a characteristic phase transition
phenomenon peculiar to bistable systems.

For $\beta, \eta$ lying far from the phase transition line $\beta_0(\eta)$,
the supernarrow peak is not seen, but there remain (histograms of Fig 10) the
much broader and less intense peaks in the SDF corresponding to fluctuations
about the stable states.  These correspond to the partial spectra of the first
term in (18) and are well described (full curves) by (19).  The characteristic
feature is that they each ({\it j} = 1, 2) display a twin-peaked structure for
a certain range of parameters.  It can be seen from (19) that such structure
should be at its most pronounced for the spectrum $Q_j(\Omega)$ when
$\vert \nu_j \vert \gg 1$: that is, for sufficiently large frequency detuning
$\vert \omega_F - \omega_0 \vert \gg \Gamma$.
Under these conditions, the peaks appear at $(\omega_F - \omega_0)
\simeq \pm \Gamma \nu_j$, and their half-width $\sim \Gamma$ is small compared
to the distance separating them.  The twin-peaked structure can be understood
intuitively in terms of the forced vibrations at frequency $\omega_F$ in a
given stable state being modulated by the relatively slow (characteristic
frequency $\sim \omega_F - \omega_0$) fluctuational vibrations about this
state.  We notice that the intensities of the peaks in a doublet differ
markedly (parametrically strongly, in the case of the small-amplitude
attractor), so that the intensity of the second peak for the small-amplitude
attractor is fairly small.

In the range of the kinetic phase transition, the partial spectra $Q_1(\Omega),
Q_2(\Omega)$ are superimposed and the supernarrow peak is also present.  Thus
there can be up to five separate peaks in the spectrum.  A multi-peaked
structure is clearly evident in the results of Fig 12, which were recorded for
a larger detuning and a smaller $\Gamma$.  A satisfactory quantitative
description of such a spectrum cannot, however, be arrived at on the basis
of (19), because it is significantly influenced by higher order terms in the
noise intensity, i.e. by vibrations at the overtones of $\Gamma u_j$, that
were ignored in the derivation of (19).  A detailed investigation of such
higher order effects is currently being planned and will be the subject of
a future paper.

\vskip 0.3truecm
\noindent
\underbar {C.  High-frequency stochastic resonance}

In searching for evidence of the predicted high frequency stochastic
resonance (HFSR) phenomenon, the circuit parameters were: 2$\Gamma$ =
0.0397; $\omega_0$ = 1.00; $\gamma$ = 0.1; $\omega_F$ = 1.07200; $\Omega$ =
1.07097; $F$ = 0.068; $A$ = 0.006.  The frequencies of the additional
weak trial force and the main periodic drive were therefore very close to
each other.  A typical SDF, measured for a 16K digitized time series in the
memory of the Nicolet 1280 data processor with input noise intensity $B$
= 0.040, is shown in Fig 13.  The central maximum is the supernarrow SDF
peak of Fig 10(b), here with its finite width clearly resolved (note the
highly expanded abscissa scale).  A delta function spike is evident, not
only at $\Omega$, but also at the mirror-reflected frequency $(2 \omega_F
- \Omega)$ just as predicted in Section IIC.

The signal strengths (integrated intensities) $S (\alpha)$, ${\cal S}
(\alpha)$,
determined from measurements of the delta spikes, are plotted (data points) as
functions of the reduced noise intensity $\alpha \propto B$ in Fig 14(a) and
(b).  It is immediately apparent that there are well-defined maxima in the
plots of $S(\alpha)$, ${\cal S} (\alpha)$, thereby confirming the occurrence
of HFSR in (3).  The signal/noise ratio $P, {\cal P}$ (i.e. the ratio of $S$,
${\cal S}$ to the \lq\lq background" SDF in the absence of the trial force)
have also been measured.  As shown in Fig 15, although the statistical quality
of the data is somewhat poorer (owing to the additional error in the background
SDF), there is no doubt that $P, {\cal P}$ each fall and rise and fall again
with increasing noise intensity.  The fall in $P, {\cal P}$ with increasing
$\alpha$ at small $\alpha$ is, of course, a feature that is familiar from
earlier calculations and experiments on conventional SR involving static
bistable attractors; the signal/noise falls initially because, for very weak
noise, the inter-attractor transitions are too rare to make significant
contributions either to the susceptibility or to the SDF, whereas the
background SDF in the denominator corresponding to fluctuations about the
attractors steadily increases with noise intensity.

The theoretical predictions, based on Eqs (18), (19), (21) (24-26) and
(28), are shown by the full curves in Figs 14 and 15.  The agreement is not
perfect but (given the problem with the determination of $\beta$: see Section
III), it is within the experimental error and may be regarded as satisfactory.
The onset of the observed rise in $S, {\cal S}, P, {\cal P}$ occurs at the
value $\alpha_0$ of noise intensity for which the width of the supernarrow
peak in the SDF becomes comparable with the frequency difference $\vert \Omega
-  \omega_F \vert$ (provided that the latter is not itself exponentially
small; cf [7b] where the position of the minimum of the SNR versus noise
intensity has been discussed for stochastic resonance in a system fluctuating
in a static bistable potential).  It is the increasing role of fluctuational
transitions that is responsible for high frequency stochastic resonance.
These results, and the good agreement obtained with the theory of Section II,
demonstrate that HFSR for periodic attractors may be perceived as a linear
response phenomenon, in very close analogy to conventional SR for a static
bistable potential [7].

An intuitive understanding of the mechanism of HFSR can be gained by
recalling that, under the conditions considered here with $\vert \Omega -
\omega_F \vert$ very small, the system responds to the trial force almost
adiabatically.  In terms of the phase diagram Fig 1, the beat envelope of the
combined main and trial periodic forces results in a slow vertical oscillation
of the operating point p.  When this is set (see line ${\rm p}^{\prime} -
{\rm p} - {\rm p}^{\prime \prime}$) to straddle the KPT line, which was the
case for present investigations, and the noise intensity is in the appropriate
range, the system will have a tendency to make inter-attractor transitions
{\it coherently}, once per half-cycle of the beat frequency.  The net effect
is to increase the modulation depth of the beat envelope of the
response, thereby amplifying its component frequencies $\Omega$ and
$\vert \Omega - 2 \omega_F \vert$.

The magnitude of the signal at $\Omega$ has been measured as a function
of distance, expressed in terms of $\beta$, from the KPT.  The result
is shown in Fig 16.  It exhibits a fast cusp-like (note the log scale)
decrease of $S$ as $\beta$ moves away from its critical value, demonstrating
that, like the associated supernarrow spectral peak (see Section IV B),
HFSR for periodic attractors has the character of a critical phenomenon,
in agreement with the theory of Section II.

\vskip 0.3truecm
\centerline {V.  CONCLUSION}

Studies of a very simple nonequilibrium bistable system - a nearly resonantly
driven nonlinear oscillator - have enable us to observe and investigate a
number of new phenomena of rather general applicability and, in doing so, to
test a theoretical approach to the calculation of transition probabilities in
noise-driven nonequilibrium systems.  The onsets of the supernarrow peak in
the spectral density of the fluctuations and the corresponding peak in the
susceptibility, and the phenomenon of high frequency
stochastic resonance, can all be viewed as examples of critical kinetic
phenomena in periodically driven systems.  They may be used, not only
to investigate the character and properties of kinetic phase transitions (as
here), but also to obtain tunable noise-induced amplification of the
signal/noise ratio and extremely narrow-band filtering and detection of
high-frequency signals.

Finally, it is interesting to note that many of the phenomena discussed
above provide illustrations of the {\it creative} role often played
by noise in nonlinear systems.  The occurrence of the extremely strong
and highly selective four-wave mixing, for example, arises because
of the noise, and does not occur in the absence of noise; the dependences
on noise intensity both of this effect, and of the other critical
phenomena discussed in the present paper, are exponentially sharp.

\vskip1cm
\centerline {ACKNOWLEDGEMENTS}

Two of us (MID and DGL) acknowledge the extremely warm hospitality of Lancaster
University, where this work was done during their visits.  The research was
supported by the Science and Engineering Research Council (UK), by
the European Community and by the Ukrainian Academy of Sciences.

\newpage
\centerline {APPENDIX}

The calculation in [3] of transition probabilities for systems driven by
Gaussian noise was based on an idea [25] due to Feynman.  His suggestion
was that there was a direct relationship between the probability density of the
paths of the noise-driven system and the noise itself.  This inter-relationship
allows us to write immediately, to logarithmic accuracy, the probability
density functional for the paths of the system, and to set up the variational
formulation giving the most probable paths for first reaching a given point
in the phase space of the system and for the transitions between the stable
states (see also Ref 2).  In the white noise case under consideration, the
\lq\lq activation energy" $R_j$ characterising the transition of the
oscillator from the stable state {\it j} to the stable state {\it i} is
given [3] by the following variational problem

$$R_j = \frac {1}{4} \eta^{2} {\rm min} \int^{\infty}_{-\infty} d\tau \left(
\frac {du}{d\tau} - \upsilon \right) \left( \frac {du^*}{d\tau} -
\upsilon^* \right)\eqno(A1)$$

$$\upsilon \,\, \equiv \,\, \upsilon (u, u^*), \qquad u(- \infty) \,\, = \,\,
u_j, \qquad u(\infty) \,\, = \,\, u_s$$

\noindent
where $\upsilon$ is defined by (6), and $u_j$ and $u_s$ are the values of the
\lq\lq slow" variable {\it u} for the initially occupied stable state and for
the saddle point respectively.  The general analysis of large occasional
fluctuations in systems driven by white noise was given by Wentzell' and
Freidlin [26].

In obtaining a variational (Euler) equation for the problem (A1), {\it u}
and $u^*$ should be varied independently.  The resulting equation can be seen
to be of the form

$$\frac {d^2u}{d\tau^2} \,\, - \,\, 2i \frac {du}{d\tau} (2 \vert u\vert^2 -1)
\,\, - \,\, \eta^{2} u\nu^2 + \sqrt {\beta} (2 \vert u \vert^2 + u^2 - 1 - i
\eta) = 0 \eqno(A2)$$

$$\nu^2 \,\, \equiv \,\, \nu^2(u, u^*) \,\, = \,\, 1 + \eta^{-2} (\vert u
\vert^2  - 1)(3 \vert u \vert^2 - 1)$$

\noindent
together with the equation for $u^*$ complex conjugate to (A2).  The
corresponding equations for $u^{\prime} \equiv {\rm Re} u$, $u^{\prime \prime}
= {\rm Im} u$ were written down explicitly in Ref [2].  An analytic solution
can be obtained [2] in some limiting cases.  In particular, in the small
$\eta$ limit, the equations describe the conservative motion of an auxiliary
system with two degrees of freedom, its coordinates being
$u^{\prime}, u^{\prime \prime}$ and its velocities $\dot u^{\prime}$ and
$\dot u^{\prime \prime}$.  The motion can be considered as the planar motion
of a particle of unit mass and unit electric charge in an electric potential

$$U (u^{\prime}, u^{\prime \prime}) = -\frac {1}{2} (\frac {dg}{du^{\prime
\prime}} + \eta u^{\prime})^2 - \frac {1}{2} (\frac {dg}{du^{\prime}}
- \eta u^{\prime \prime})^2\eqno (A3)$$

\noindent
where

$$g(u^{\prime}, u^{\prime \prime}) = \frac {1}{4} (u^{\prime 2}
+ u^{\prime \prime 2} - 1)^2 - u^{\prime} \sqrt {\beta}$$

\noindent
and a magnetic field ${\cal H} = [4 (u^{\prime 2} + u^{\prime \prime 2}) - 2]$
normal to the plane.  The potential $U (u^{\prime}, u^{\prime \prime})$ is
shown in Fig 17.  It has three maxima of equal height (=0).  They
correspond respectively to the stable states and the unstable stationary
state of the system.  Note that we are dealing with the {\it auxiliary}
system.  Thus, not minima, but maxima of the potential $U (u^{\prime},
u^{\prime \prime})$ correspond to the stable states of the initial system
(a circumstance typical of the instanton formulation that we are using).  The
problem (A1) amounts to finding a path that starts on one of the outer
maxima and arrives at the maximum corresponding to the saddle point.

The numerical solution in the general case of arbitrary values of
($\beta, \eta$) can be simplified by the following procedure (cf [27]).  Near
the stable state {\it j}, when $\vert u - u_j \vert \ll 1$, Eq (A2) can be
linearized in $u - u_j, u^* -u_j^*$.  The solution can then be sought in
the form

$$u(\tau) \,- \,u_j \,\, = \,\, \sum \limits^{}_{s} A^{(s)}_j {\rm exp}
(\lambda^{(s)}_j \tau) \eqno(A4)$$

$$u^* (\tau) \, - \, u_j^* \,\, = \,\, \sum \limits^{}_{s}B^{(s)}_j {\rm exp}
(\lambda^{(s)}_j  \tau)$$

\noindent
for $\tau \to - \infty$, with

$$B^{(s)}_j \,\, = \,\, A^{(s)}_j \,\, \frac {\lambda^{(s)^2}_j -2i \lambda_j
^{(s)} (2 \vert u_j \vert^2 - 1) - (\eta^{2} + 5 \vert u_j \vert^4 -4 \vert
u_j \vert^2 + 1)}{2u_j^2 (2\vert u_j \vert^2 - 1 - i \eta)}\eqno(A5)$$

\noindent
The resultant characteristic equation gives four values for the increment
$\lambda$

$$\lambda^{(1, 2)}_j \,\, = \,\, \alpha^{(1, 2)}_j, \quad \lambda^{(3)}_j
= - \alpha^{(1)}_j, \quad \lambda^{(4)}_j = - \alpha^{(2)}_j \eqno(A6)$$

\noindent
where $\alpha_{j}^{(1, 2)}$ are the roots of the characteristic equation
for the motion of the oscillator in the vicinity of the state {\it j} in the
absence of the random force (the latter being described by the linearized
equation $du/ d\tau = \upsilon$),

$$\alpha^{(1, 2)}_j \,\, =\,\,- \eta [1 \pm i (\nu^2_j - 1)
^{\frac {1}{2}}], \quad \nu^2_j \, \, \equiv \,\, \nu^2(u_j, u_j^*) \eqno(A7)$$

\noindent
For the stable state, Re $\alpha_j^{(1,2)} < 0$; note that this implies that
$\nu^2_j > 0$.

It is evident from (A4)-(A7) that the coefficients $A^{(1, 2)}_j$ (and thus
$B_j^{(1, 2)}$) in (A4) should be set equal to zero; otherwise, the path
$u(\tau)$ will not approach $u_j$ as $\tau \to - \infty$.  At this point we
have arrived at two independent parameters in (A4): $A^{(3)}_j$ and
$A^{(4)}_j$.
However, it is their {\it ratio} which determines the direction in which the
system will move along the extreme path (A2); accordingly, it is this ratio
that should be determined from the boundary conditions $u(0) = u_s$ (A1).  We
thus obtain an implicit equation for the single quantity $A^{(3)}_j/A^{(4)}_j$.
It can in principle be solved numerically by a shooting method (cf [28]).
This method works most effectively for small damping, $\eta \ll 1$, when the
optimal path $u^{\prime} (\tau), u^{\prime \prime} (\tau)$ is a spiral.

Here, we adopt a relaxation method [29].  The differential equations are
cast in the form of nearest neighbour difference equations.  The boundary
values of $u, u^*$ were chosen at $u_j, u_s$.  (The results were the same
to the adopted accuracy when $i_{ij}$ was taken instead of $u_s$, i.e.
the optimal path for the escape from the state $j$ was supposed to start
from $u_j$ and to arrive at the other stable position $u_{3-j}$;
moreover, it turned out that the path found in this way passed through,
or fairly close to, the saddle point).  A guess at the solution was tried,
and then it was successively improved by assuming, at each step, that
the true solution was close to the current one and linearising about the
latter.  Even for fairly different initial guesses, the same final solution
$u(t)$ was obtained, implying that the method is reliable.  The same
results were obtained for different integration times, which were always very
large compared to the characteristic dimensionless times $\sim 1, \eta^{-1}$,
in (A2).

Solutions were used to compute the action integral (A1).  The results are
summarised in Fig 17.  The optimal path is indicated by the $\diamond$
points, which are separated by equal intervals of time.  The motion
is naturally at its slowest (points closest together) on the maxima of
the potential; because of the effect of the \lq\lq magnetic field" ${\cal H}$,
it does not pass {\it exactly} along the ridges.  The advantage of the
relaxation method is that it is fairly fast and convenient.  On the other hand,
for small $\eta$ where the optimal path is a small-step spiral, it is less
reliable than the method based on solving Eqs (A4)-(A7) which was used in [28].

When, in addition to the strong field $F \cos \omega_Ft$, the
oscillator is also driven by a weak force $A \exp (-i \Omega t)$,
the expressions for $R_j$ will change.  The additions to $R_j$ can readily be
found when $\vert \Omega - \omega_F \vert \ll \Gamma$, because the
characteristic time of the motion along the extreme path described by (A2) is
$\Gamma^{-1}$ ({\it i.e.} $\eta^{-1}$ in dimensionless units of $\tau$); thus,
when $\vert \Omega - \omega_F \vert \gg \Gamma$, the weak field is
not changed while the system is moving along the path.  It is evident from
(22) that the functional which should be minimised to give $R_j$ in the
presence of the field $A$ is given by (A1) with $\upsilon^*$ having
been replaced by $\upsilon^* + i \tilde {A} (\tau)$.  To first
order in $A$, the change in $R_j$ is thus of the form

$$\delta R_j (\tau) \,\, = \,\, \mu_j \tilde {A} (\tau), \quad
\mu_j \,\, = \,\, - \frac {i}{4} \eta^{2} \int^0_{-\infty} d\tau
(\frac {du}{d\tau} - \upsilon )\eqno(A8)$$

\noindent
where the integral giving $\mu_j$ is calculated along the extreme path for
$A = 0$.

The expression for $\mu_j$ can be substantially simplified if one notices
that the activation energy $R_j$ (A1) is unchanged when $\beta^{\frac {1}{2}}$
is replaced by $\beta^{\frac {1}{2}} {\rm exp} (i \psi)$ in (6) and (A1)
respectively, where $\psi$ is arbitrary: such a replacement corresponds
simply to a shift of the time origin in (5) by $\psi / \omega_F$, which
should not influence stationary characteristics of the oscillator such as
$R_j$.  By differentiating $R_j$ with respect to $\psi$ for $\psi \to 0$,
one finds immediately that ${\rm Im}\mu_j = 0$, and it is then easy to see that

$$\mu_j \,\, = \,\, \sqrt {\beta} \,\, \partial R_j/\partial \beta \eqno(A9)$$

\newpage
\centerline {REFERENCES}

\begin{itemize}
\item[1.]  H. A. Kramers, {\it Physica} {\bf 7}, 284 (1940).

\item[2.] M.I. Dykman and K. Lindenberg, in {\it Some Problems of
Statistical Physics}, ed G Weiss (SIAM, Philadelphia, 1993).

\item[3.]  (a) M. I. Dykman and M. A. Krivoglaz, {\it Zh. Eksp. Teor. Fiz.}
{\bf 77}, 60 (1979) [{\it Sov. Phys. J.E.T.P.} {\bf 50}, 30 (1979)]; and (b)
in {\it Soviet Physics Reviews}, ed. I. M. Khalatnikov (Harwood Academic,
New York, 1984), vol. 5, p 265.

\item[4.] M. I. Dykman, R. Mannella, P. V. E. McClintock, F. Moss and S. M.
Soskin, {\it Phys. Rev. A} {\bf 37}, 1303 (1988).

\item[5.] (a) M. I. Dykman, P. V. E. McClintock, R. Mannella and N. G. Stocks,
Soviet Phys.  {\it JETP Lett.} {\bf 52}, 141 (1990); M. I. Dykman, R. Mannella,
P. V. E. McClintock and N. G. Stocks, {\it Phys. Rev. Lett.} {\bf 65}, 2606
(1990); and (b) {\it ibid} {\bf 68}, 2985 (1992).

\item[6.] R. Benzi, A. Sutera and A. Vulpiani, {\it J. Phys. A} {\bf 14}, L453
(1981); C. Nicolis, {\it Tellus} {\bf 34}, 1 (1982); R. Benzi, G. Parisi,
A. Sutera and A. Vulpiani, {\it Tellus} {\bf 34}, 10 (1982).

\item[7.] B. McNamara, K. Wiesenfeld and R. Roy, {\it Phys. Rev. Lett.}
{\bf 60}, 2626 (1988); R. F. Fox, {\it Phys. Rev. A} {\bf 39}, 4148 (1989);
B. McNamara and K. Wiesenfeld, {\it Phys. Rev. A} {\bf 39}, 4854 (1989);
Hu Gang, G. Nicolis and C. Nicolis, {\it Phys. Rev. A} {\bf 42}, 2030 (1990);
M. I. Dykman, G. P. Golubev, D. G. Luchinsky, A. L. Velikovich and S. V.
Tsuprikov, {\it Soviet Phys. JETP Lett} {\bf 53}, 193 (1991).

\item[8.] For a fuller bibliography and reviews of the rapidly
expanding field of stochastic resonance, see the special issue of {\it J.
Stat. Phys.} vol 70, Nos. 1/2 (1993) and the papers and references therein.

\item[9.] H. M. Gibbs, {\it Optical Bistability: Controlling Light with
Light}, (Academic Press, New York, 1985).

\item[10.] L. D. Landau and E. M. Lifshitz, {\it Mechanics} (Pergamon, London,
1976).

\item[11.] G. Gabrielse, H. Dehmelt and W. Kells, {\it Phys. Rev. Lett.}
{\bf 54}, 537 (1985).

\item[12.] M. I. Dykman, M. A. Krivoglaz and S. M. Soskin, in {\it Noise in
Nonlinear Dynamical Systems}, ed. F. Moss and P. V. E. McClintock
(Cambridge University Press, 1989), vol 2, p 347.

\item[13.] M. I. Dykman, R. Mannella, P. V. E. McClintock and N. G. Stocks,
{\it Phys. Rev. Lett.} {\bf 65}, 48 (1990).

\item[14.] B. A. Huberman and J. P. Crutchfield, {\it Phys. Rev. Lett.}
{\bf 43}, 1745 (1979).

\item[15.] See, for example, J. M. T. Thompson and H. B. Stewart,
{\it Nonlinear Dynamics and Chaos} (Wiley, New York, 1987) and references
therein.

\item[16.] F. T. Arecchi, R. Badii and A. Politi, {\it Phys Rev A}
{\bf 32}, 402 (1985).

\item[17.] N. N. Bogolyubov and Yu A Mitropolsky, {\it Asymptotic Methods in
the Theory of Nonlinear Oscillators}, (Gordon and Breach, New York, 1961).

\item[18.] M. I. Dykman and V. N. Smelyanski, {\it Phys. Rev. A} {\bf 41},
3090 (1990).

\item[19.]  R. Bonifacio and L. A. Lugiato, {\it Phys. Rev. Lett.} {\bf 40},
1023 (1978); L. A. Lugiato, {\it Progr. Optics} {\bf 21}, 69 (1984).

\item[20.] L. D. Landau and E. M. Lifshitz, {\it Statistical Physics}
(Pergamon, New York, 1980), 3rd edn., part 1.

\item[21.] Stochastic resonance was originally considered (see Ref 6) just as
a periodic-field-induced periodic redistribution over the stable states.

\item[22.] Y. R. Shen, {\it The Principles of Nonlinear Optics} (Wiley,
New York, 1984).

\item[23.] L. Fronzoni in {\it Noise in Nonlinear Dynamical Systems},
ed. F. Moss and P. V. E. McClintock (Cambridge University Press, 1989),
vol 3, p 222; and P. V. E. McClintock and F. Moss in {\it loc. cit.}
p 243.

\item[24.] P. H\"anggi, T. J. Mroczkowski, F. Moss and P. V. E. McClintock,
{\it Phys. Rev. A.} {\bf 32}, 695 (1985).

\item[25.] R. P. Feynman and A. R. Hibbs, {\it Quantum Mechanics and Path
Integrals}, (McGraw-Hill, New York, 1965).

\item[26.] A.D. Wentzel' and M.I. Freidlin, {\it Russ. Math. Surveys}
{\bf 25}, 1 (1970); M.I. Freidlin and A.D. Wentzel', {\it Random
Perturbations of Dynamical Systems} (Springer, New York, 1984).

\item[27.] D. Ludwig, {\it SIAM Rev.} {\bf 17}, 605 (1975).

\item[28.] V. A. Chinarov, M. I. Dykman and V. N. Smelyanski, {\it Phys Rev E}
{\bf 47}, 2448 (1993).

\item[29.] W. H. Press, B. P. Flannery, S. A. Teukolsky and W. T.
Vetterling, {\it Numerical Recipes} (Cambridge University Press, 1986).
\end{itemize}

\newpage
\centerline {FIGURE CAPTIONS}

\begin{itemize}
\item[1.] Phase diagram for the system (3) in terms of the reduced
parameters (7).  Within the approximately triangular region enclosed
by the full lines, the system is bistable, with two possible stable
limit cycles of different amplitude and phase relative to the
periodic driving force.  Outside this region, the system is monostable.
The dashed line represents the calculated position of the kinetic phase
transition.  The cuts a-a$^{\prime}$ and p$^{\prime}$-p-p$^{\prime \prime}$
are discussed in the text.

\item[2.] The evolution of the basins of attraction for (3), computed
for $\eta^2$ = 0.072 in Poincar\'e section with the same phase as that of the
driving field.  The white regions show the attracting basin of the large
amplitude attractor, and the black regions show the basin for the small one;
each attractor is indicated by a $\bullet$.  The values of $\beta$,
left-to-right, from top-to-bottom, were: 0.0709; 0.0811; 0.0913; 0.102; 0.112;
0.122; 0.132; 0.143; 0.153; and 0.163.

\item[3.] Calculated activation energies $R_i$ for transitions between
the coexisting periodic attractors of (3), as functions of $\beta$ for: (a)
$\eta^{2}$ = 0.033; (b) $\eta^{2}$ = 0.100; (c) $\eta^{2}$ = 0.200; (d)
$\eta^{2}$ = 0.333.  The circle data points were obtained by minimisation
of the action energy integral $R_j$ in Eq (A1); the curves are derived
from analytic expansions [3] around the relevant birfurcation points.
In each case, the falling data and curve represent $R_1$ and the rising data
and curve represent $R_2$.

\item[4.] Block diagram of the analogue electronic circuit model
of Eq (3).

\item[5.] (a) Variations of the coordinate $q (t)$ with time $t$, measured for
the electronic circuit model of (3) within the bistable regime in the absence
of noise with $\beta$ = 0.0607, $\eta^2$ = 0.033.  Digitized time series for
the small and large amplitude attractors are shown superimposed; note the
phase difference between them.  In (b) the instantaneous energy $E$ of the
system is plotted as a function of $q$: the lower curve is for the small
amplitude attractor and the upper curve is for the larger one.

\item[6.] Comparisons of the calculated region of bistability (between
the full curves: cf Fig 1) with that measured for the electronic circuit model
of (3) (square data points), and between the calculated (dashed line) and
measured (crosses) kinetic phase transition line.

\item[7.] (a) Variation of the coordinate $q(t)$ with time $t$, measured
for the electronic circuit model of (3) in the bistable regime of the
presence of noise with $\beta$ = 0.0607, $\eta^2$ = 0.033, showing a transition
between the attractors (a rare event).  (b) The corresponding variation of the
instantaneous energy $E$ with the coordinate $q$ (cf Fig 5 in the absence of
noise).

\item[8.] (a) Dependence of the apparent mean first passage time on the
position chosen for the upper criterion level for crossings: the energy of the
larger attractor for $q$ = 0 in the absence of noise is shown by the dashed
line.  In practice, all measurements were made in the plateau region.  (b)
Logarithmic plots of the MFPT between the attractors measured for the analogue
electronic circuit model of (3) as a function of reciprocal noise intensity
$\alpha^{-1}$, for $\eta^2$ = 0.033: + from small amplitude attractor
with $\beta$ = 0.0888; $\Box$ from large attractor with $\beta$ = 0.047;
$\bigcirc$ from small attractor with $\beta$ = 0.0734.  The fact that the data
lie on straight lines demonstrates the \lq\lq thermal activation" character of
the transition mechanism.

\item[9.] Values of the activation energies $R_i$ measured (crosses) as
functions of
$\beta$ for the analogue electronic circuit model of (3) with $\eta^2$ =
0.033: (a) for $R_1$; and (b) for $R_2$.  The theoretical curves are
the same as those of Fig 3(a).

\item[10.] Spectral densities $Q (\omega)$ of fluctuations measured
(histograms) for the analogue electronic circuit model of (3) with $\eta$ =
0.219, $\alpha$ = 8.69 $\times$ 10$^{-2}$ for: (a) $\beta$ = 0.048; (b)
$\beta$ = 0.078; (c) $\beta$ = 0.150.  The full curve represents the theory.

\item[11.] Variation of the intensity $I$ of the supernarrow spectral
peak with distance from the kinetic phase transition line, measured
as a function of $\beta$ for the analogue electronic circuit model
of (3) for $\eta$ = 0.219, $\alpha$ = 8.69 x 10$^{-2}$.  The
squares represent direct measurements; the crosses are derived
from (21), based on measured transition rates.  The full lines also represent
(21), but for ln$(w_1 w_2)$ given by (31) with measured
$R_1^{\prime}, R_2^{\prime}$.

\item[12.] An example of the kind of multipeaked spectral density of
fluctuations $Q(\omega)$, measured in the range of the kinetic phase
transition for the analogue electronic model of (3), with relatively
large frequency detuning and small damping.  The parameters were: $\eta$ =
0.055; $\beta$ = 0.0303; $\Gamma$ = 0.0073; $\alpha$ = 0.238.  Five
spectral peaks - the supernarrow peak and four peaks of the partial
spectra - are clearly resolved.

\item[13.] Spectral density $Q (\omega)$ of the fluctuations of (3) as
a function of frequency $\omega$ with an additional weak periodic (trial)
force $A \cos (\Omega t + \phi)$ added on the right hand side, measured for
the analogue electronic circuit model.  The contents of each FFT memory address
are shown as a separate data point on a highly expanded abscissa (unscaled
experimental unit).  A smooth curve has been drawn through the background
spectrum (the supernarrow peak, which has its maximum at $\omega_F$) as a
guide to the eye; vertical lines indicate the delta spikes resulting from
the trial force.

\item[14.] The intensities $S$ and ${\cal S}$ of the $\delta$-shaped peaks
in the SDF of the analogue electronic model of (3) (data points) with $\beta$
= 0.103, $\eta$ = 0.266 induced by a weak trial force $A \cos \Omega t$,
plotted as a function of noise intensity $\alpha$, compared with theory
(full curves): (a) at the trial force frequency $\Omega$; (b) at the
mirror-reflected frequency (2$\omega_F - \Omega$).

\item[15.] The signal/noise ratios $P$ and ${\cal P}$ of the responses at
$\beta$ = 0.814, $\eta$ = 0.236 to the trial force at frequencies $\Omega$
(circle data and associated curve) and ($2 \omega_F - \Omega)$ (squares),
measured as functions of noise intensity $\alpha$ for the analogue
electronic circuit model of (3).  The curves represent the theory.
For noise intensities near those of the maxima in $P (\alpha)$
${\cal P} (\alpha)$, the asymptotic theory is only qualitative and so the
curves are shown dotted.

\item[16.] Dependence of the logarithm of the intensity $S$ of the spectral
peak induced by the trial force for the analogue electronic circuit model of
(3) at a fixed noise intensity $\alpha$ = 0.076, plotted as a function of
$\beta$, varying across the KPT line.

\item[17.] The three-humped potential $U (u^{\prime}, u^{\prime \prime})$
of Eq (A3) for the auxiliary system, portrayed both as a three-dimensional
surface and, below it, in the form of a contour plot; the contour
altitudes are tabulated on the right hand side.  (Note, however, that
the left hand potential maximum is so shallow that it is barely visible in
the contour plot).  The optimal path of the escape from the focus-1
(the small amplitude attractor) is shown by the line of $\diamond$ points.
It goes from this focus (the right hand potential maximum) to the saddle
point (central maximum).  On its way from the saddle point to the focus-2
(the left-hand potential maximum corresponding to the large amplitude
attractor) the system moves, with overwhelming probability, along the
noise-free path.  The $\diamond$ points are equally spaced in time, so that the
speed with which the system is travelling along different elements of the path
may be inferred from the density of the points; it moves most slowly at the
three potential maxima.  The values of the dimensionless parameters for the
plot were $\eta^2$ = 0.072, $\beta$ = 0.104.
\end{itemize}
\end{document}